\documentstyle[12pt]{article}

    \newlength{\absize}
\begin{document}
  \newcommand\Bkg{$B \rightarrow K^\ast \; \gamma\,$ decay}
  \newcommand\Brg{$B \rightarrow \rho \; \gamma\, $ decay}
  \newcommand\bsg{$b \rightarrow s \; \gamma\,$ decay}
  \newcommand\bdg{$b \rightarrow d \; \gamma \,$ decay}
  \newcommand\Bdg{$B \rightarrow  X_d \; \gamma \,$ decay}
  \newcommand\gev{{\rm GeV }}
  \newcommand\mev{{\rm MeV }}
  \newcommand\vub{$V_{ub} $}
  \newcommand\vcb{$V_{cb} $}
  \newcommand\vtb{$V_{tb} $}
  \newcommand\vud{$V_{ud}^\ast $}
  \newcommand\vcd{$V_{cd}^\ast $}
  \newcommand\vtd{$V_{td}^\ast $}
  \newcommand\vuds{$V_{u(d,s)}^\ast $}
  \newcommand\vcds{$V_{c(d,s)}^\ast $}
  \newcommand\vtds{$V_{t(d,s)}^\ast $}
  \newcommand\beq{\begin{equation}}
  \newcommand\eeq{\end{equation}}
  \newcommand{\preprint}[1]{%
    \begin{flushright}
      \setlength{\baselineskip}{3ex} #1
    \end{flushright}}
  \renewcommand{\title}[1]{%
    \begin{center}
      \LARGE #1
    \end{center}\par}
  \renewcommand{\author}[1]{%
    \vspace{2ex}
    {\normalsize
     \begin{center}
       \setlength{\baselineskip}{3ex} #1 \par
     \end{center}}}
  \renewcommand{\thanks}[1]{\footnote{#1}}
  \renewcommand{\abstract}[1]{%
    \vspace{2ex}
    \normalsize
    \begin{center}
      \centerline{\bf Abstract}\par
      \vspace{2ex}
      \parbox{\absize}{#1\setlength{\baselineskip}{2.5ex}\par}
    \end{center}}
  \begin{titlepage}
  \preprint{DSF-T-54-97\\ITP-SB-97-69\\HUTP-97/A058\\Revised}
 
\vspace{3mm}

  \title{
  Light masses in short distance\\ penguin loops}
  \vskip 0.3cm

\begin{center}
{\large Mario Abud$\; {}^{a,b,}$\footnote{\sl e-mail: mario.abud@napoli.infn.it},
Giulia Ricciardi$\; {}^{b,c,}$\footnote{\sl e-mail: ricciardi@huhepu.harvard.edu},
George Sterman$\; {}^{d,}$\footnote{\sl e-mail: sterman@insti.physics.sunysb.edu}}

\vspace{3mm}

$^a${\sl Dipartimento di Scienze Fisiche,
          Universit\`a  degli Studi di Napoli\\
Mostra d' Oltremare, Pad 19, I-80125 Napoli, Italy}\\

\vspace{2mm}

$^b${\sl INFN (Sezione di Napoli)\\
Mostra d' Oltremare, Pad 19, I-80125 Napoli, Italy}

\vspace{2mm}

$^c${\sl Lyman Laboratories,  Harvard University, Cambridge, MA 02138, USA}

\vspace{2mm}

$^d${\sl Institute for Theoretical Physics,
          SUNY Stony Brook\\ Stony Brook, NY 11794-3840, USA}
\end{center}

\vspace{4mm}

\abstract{
Penguin
diagrams for decays such as $b\rightarrow (s,d)\gamma$ involve
 virtual loops of $u$ or other light quarks.  
Logarithms of the virtual quark mass 
could, in principle, influence the
phenomenological analysis of the decay. It is thus important
to study these logarithms to all orders in QCD perturbation
theory.
In this paper we show  that,  at arbitrary order, 
the matrix elements of  operators in
  the effective hamiltonian
  contributing to $b\rightarrow s\gamma$ are
  finite for the limit of $m_u \rightarrow 0$ 
in penguin loops.}

  \end{titlepage}
  \vskip2truecm 

\setcounter{footnote}{0}
  \section{Introduction}

  \label{sec:intro}

In the spectator-quark model,
    the inclusive weak decays of the $B$-meson can be pictured as 
QCD corrected $b$-quark decays.
  Since in perturbative QCD, virtual light quarks
  appear in loop diagrams,
 their presence might suggest
an enhancement of the rate through
terms involving powers of $\ln m_{loop}$
(we denote by $m_{loop}$ 
the mass of any light quark circulating in the
 penguin loops).   The possible effects of such
 terms ``long-distance" contributions were
  explored, for example, in Ref. \cite{Deshpande:LDPenguins}.
 On the other hand, the finiteness for
$m_{loop} \rightarrow 0$ has been invoked in other papers, such as
\cite{ali:bdg_1,Soares,ali:bdg_2}.
  In this paper, we address this problem
  in the context of effective field theory.
Our analysis is perturbative and therefore  relates to current,
rather than constituent quark masses \cite{soni-vtd}.

  We refer to a particular flavor changing neutral current process,
  $b \rightarrow s \gamma $ decay, but our results can easily be extended
  to other interesting FCNC processes, 
such as $b \rightarrow d\gamma$.
  We argue that, to any order in perturbative QCD,
  the limit $m_{loop} \rightarrow 0$ in gauge invariant penguin amplitudes is
  finite, and the presence of virtual light
  quarks in these internal loops  does not result in
  logarithmic divergences as $m_{loop} \rightarrow 0$.
To be specific we argue that powers of $\ln m_{loop}$ are always accompanied
  by positive powers of $m_{loop}$ \cite{GREUB}.
For example, in $b \rightarrow (s,d) \gamma $ decays there is 
a contribution from  penguin loops with a virtual $u$ quark 
($m_{loop}=m_u$), but the amplitude 
does not diverge as $m_u \rightarrow 0$.
For definiteness, we discuss the exclusive $b\rightarrow s\gamma$ final state. 

Of course, penguin loops are not the only
source of dependence on light quark masses.
In the partonic $b\rightarrow s\gamma$ amplitude, 
subdiagrams
involving gluons (photons) attached to the  outgoing $s$-quark 
as well as soft gluons attaching the $b$-quark with the 
$s$-quark jet, are collinear divergent in perturbation theory in general.
In practical calculations, these regions give logarithms of $m_s$
that are not suppressed by powers of $m_s$.
In addition, the amplitude for any
exclusive final state contains purely infrared divergences associated with
the masslessness of the gluon.
These kinds of divergence have already been 
treated in the literature 
(see for instance \cite{GREUB,ali:radiative,Kapustinetal}).  
  As part of
our analysis on penguin loops, we shall show that
no additional logarithms arise
from light quark  loops collinear to the 
outgoing photon
in the particular case of 
the two-body $b\rightarrow s\gamma$ decay amplitude.  
  We derive our results by building the effective hamiltonian step by step,
showing at each stage which kinds of dependence on light masses are to be expected,
 and by applying analyticity arguments and IR power
  counting techniques developed in perturbative QCD \cite{sterman}.

  \section{The Effective Field Theory}
   \label{short-distance}

  The effective field theory for $B$-decays
  describes the physics at the scale $\mu \simeq m_b
  \ll m_W$, after the top quark and the $W$ boson
  have been integrated out. The
  effective hamiltonian 
at the lowest order in $\alpha_{em}$
is defined by a sum of local operators whose matrix
  elements between initial and final states reproduce the amplitude at
  low energies
  \begin{equation}
  \label{sd-eff-ham}H_{eff}=\frac{G_F}{\sqrt{2}}{\cal V}_{CKM}\sum_iC_i(\mu
  )O_i(\mu )\, ,
  \end{equation}
  where ${\cal V}_{CKM}$ denotes the appropriate factor (typically quadratic
  in CKM matrix elements).
  The utility of the effective hamiltonian formalism is that it
  separates short distance contributions, described by the coefficients,
  which can be calculated perturbatively,
  from  long distance contributions,
  incorporated in the matrix elements of the local operators.

  We can summarize the steps to build the effective hamiltonian
  for $b \rightarrow s \gamma$ as:
  (i) calculating
  the coefficients
   by  matching the full theory onto the effective theory at
  high scale;
  (ii) evolving the coefficients
   down to the scale $O(m_b)$ by the renormalization group (RG) equations;
  (iii) evaluating the matrix elements of the operators.
In the same spirit of Ref.~\cite{giulia:bdg},
we will analyze the role of the light masses during these three steps.

  \subsection{Matching}

  The first step consists in matching the effective theory onto the full
  theory. To match means to extract the coefficients $C_i$
   by comparing the amplitude in the full theory and
  in the effective theory at the same order in $\alpha_s$.
  At the matching scale,  all IR
  behavior cancels and logarithms
   of light masses are
  then eliminated in the coefficients.

  As an  example,
  we shall discuss weak penguin diagrams in $b\rightarrow s\gamma$, like the one
  shown in Fig.\ 1.
   The effective flavor changing gauge invariant couplings of the photon
  are of the type $F_{\mu \nu } \partial^\nu
  \overline{s}_L\gamma ^\mu
   b_L$  and $\overline{s}_L F_{\mu \nu }\sigma^{\mu\nu} b_R$.
  In a renormalizable theory like the standard model,
  the corresponding amplitudes must be UV finite, because
  the above operators have dimension higher than 4
   and cannot arise as counterterms. The
  amplitude of the full theory, to which the effective theory is matched,
   can be calculated by expanding in the ratio ${q}/{m_W}$, where $q$
   is the photon momentum.
  This expansion can introduce infrared
  divergences.
  Since the external photon is taken on shell,
  the mass of the quark in the loop of Fig.\ 1 acts as an IR regulator.

  The expansion in photon momentum results in terms of the form
  $F_1\,\,(q^2\gamma ^\mu$ $-q^\mu \gamma \cdot q)$
  and $F_2\,\, i\sigma ^{\mu \nu }q_\nu $.
  The form factor $F_1$ includes an IR-sensitive term
  $\frac 23({x_i-1})\ln x_i$,
  where $x_{i}={m_{i}^2}/{m_W^2}$, with $m_i$ the mass of
  quark $i$, circulating in the loop, while in
  $F_2$ $\ln x_i$ appears only multiplied by powers of $x_i$
  (see, for instance,
  Appendix B of Ref~\cite{InamiLim}).\footnote{Note, however, that
  the $F_1$ term decouples from an on-shell photon with physical
  polarization.}

  In the effective hamiltonian,
  the local operators contain only light quark
  fields; the heavy quark fields have been integrated out.
  In the corresponding diagrams
  for $b\rightarrow s\gamma$,
  there is an analogous term of the
  type $\frac 23\ln ({m_i}^2/\mu^2)$, where $\mu $ is
  introduced to fix
  the scale where the operators are renormalized,
  through the introduction
  of a counterterm proportional to
   $F_{\mu \nu } \partial^\nu
  \overline{s}_L\gamma ^\mu b_L$.
   The coefficients of the effective
  hamiltonian are found precisely from the difference of the diagrams
  of the full theory and the corresponding diagrams in the effective theory.
  If the internal quark is heavy,
  logarithms of heavy masses will be included in the matching coefficients of
  the hamiltonian. If the internal quark is light, $m_i=m_c$ or $m_u$, by
  performing the matching at $\mu =m_W$ we have an exact
  cancellation of the
  term $\frac 23\ln x_i$ in the coefficients.
  In other words, at the matching scale all IR
  behavior cancels, and the coefficients are manifestly finite for $
  m_i\rightarrow 0$.

  \subsection{RG rescaling}

  Perturbative QCD corrections
  introduce logarithms 
  $\alpha _s^n(\mu ){\ln {}^m}(\mu /m_W)$, with $m\le n$.
  The RG rescales the coefficients of the effective hamiltonian to scales
  lower than the matching scale $m_W$,
   and resums such logarithms. After the RG group rescaling,
   we are left with a residual dependence on the scale $\mu $,
  due to the truncation of the perturbative series.
  In $B$-decays, the first
  threshold is at $\mu=m_b$, and we can stop rescaling at that point;
  in $D$ and $K$ decays one can do a new matching and then use the RG once
  again to go to a still lower scale. In any case, if we are to work with
  perturbation theory, the final scale must be greater than 1
  GeV or so. It is self-evident that this step cannot introduce logarithms of
  light masses.

  \subsection{Matrix elements}

  The final step consists in calculating the matrix elements of the operators
  in the low energy theory.   
In general,  matrix elements include
  loops of virtual light quarks. We will study the limit
  where the penguin loop quark mass $m_{loop}$
goes to zero. We want to show that for the
  specially interesting case of $b\rightarrow s\gamma $ this limit also does
  not introduce IR divergences. This implies that, in the final result, any
term of the type $\ln m_{loop}$
 will always appear multiplied by powers of $m_{loop}$
 (that is, as $m_{loop}^a\ln^b m_{loop}$ with $a > 0$).
  A number of explicit calculations 
corroborate this expectation;
  see particularly Ref.~\cite{GREUB}, where
  several  matrix elements at two loops are
  calculated.

  Let us consider the Feynman diagrams that describe the matrix elements in
  the effective theory at arbitrary order in QCD perturbation theory. An
  arbitrary Feynman diagram with $N$ lines may be written in terms of Feynman parameters as
  \begin{eqnarray}
  & & G = \prod_{lines \,\, i} \int^1_0 d\alpha_i\; \delta\left(\sum_i
  \alpha_i-1\right)
  \prod_{loops \,\,r}
  \int d^4 k_r\; D(\alpha_i,k_r,p_s)^{-N}
  F(\alpha_i,k_r,p_s) \nonumber \\
  & &D(\alpha_i,l_j,p_s) =
  \sum_j \alpha_j \left(l^2_j(p_r,k_s) -m_j^2\right)+i \epsilon\, ,
  \end{eqnarray}
  where $l_j$ is the momentum of the $j$th line and $\alpha_j$ its Feynman
  parameter ($l_j$ is 
linear in the loop momenta $k_r$ and 
in the external momenta $p_s$).
  The function $F$ contains overall factors that do not enter the arguments
    at this point.
  This integral can be viewed as a contour integral in a
  multidimensional complex $\alpha_j,k_r$ space. In order to find possible
  logarithms, we have to look for the regions of non-analyticity of the
  integral. The points of the contour where $D$ vanishes are called ``singular
  points'' (SP's), and possible singularities of this integral must arise from
  zeros of the denominator $D$. In fact, only certain SP's, referred to as
  pinch SP's, give singularities that cannot be avoided by deforming the
  integration contours. Necessary conditions for a pinch SP are given by
  the so-called Landau equations~\cite{landau}.

  With each SP is associated a reduced diagram, constructed from the complete
  graph by simply contracting all lines which are off-shell at the SP. The
  reduced diagrams of pinch SP's have a direct physical
interpretation~\cite{sterman,coleman}.
  They can be interpreted as a
  picture of a classical, energy- and momentum-conserving process occurring in
  space-time, with all internal particles real, on the mass-shell, and moving
  forward in time. We may turn this interpretation around, in order to
  identify SP's that are pinched. We select the reduced diagrams associated with
  an arbitrary Feynman graph that admit such a physical interpretation; these
  diagrams identify pinch SP's.

  Once we have all the reduced diagrams relevant to a particular Feynman
  graph, we know its sources of non-analyticity. At this stage, it becomes
  important to have criteria for determining which pinch SP's may actually
  introduce infrared divergences in the diagram. The presence of
  pinch SP's reveals the presence of non-analytic terms, such as logarithms of
  light masses. If these logarithms are suppressed by powers of the light
  masses themselves, however, the corresponding amplitude will not diverge
in the zero-mass limit.

  In order to identify possible IR divergences we use IR power counting. An
  obvious complication for IR power counting in Minkowski space is that $k^2=0$
  does not imply $k=0$, so that a naive dimensional counting will not necessarily
  express the real behavior of the integral in the IR limit. A method for
  dealing with this problem is to change variables, and approximate the
  integral near each pinch SP, so that every denominator is a homogeneous
  function of a set of variables that vanish there~\cite{sterman,stbook}. This
  integral will be referred to as the ``homogeneous'' integral and these
  variables as the ``normal'' variables; the remaining variables, called
  ``intrinsic'', parametrize the relevant surface of SP's, and do not
  contribute directly to the singular behavior. The IR behavior of the
  homogeneous
  integral will be determined by dimensional power counting involving only the
  normal variables. 
In the following section, we apply the power counting procedure just sketched 
to $b\rightarrow s\gamma$, and verify that the amplitude 
is free of unsuppressed logarithms of $m_{loop}$ for $m_{loop}\rightarrow 0$.

  \section{The decay $b \rightarrow s \gamma$}

  The decay $b \rightarrow s \gamma$ has been extensively studied in the
  framework of effective field theory~\cite{bsg:inclusive}. Beyond leading
  order, the matrix elements of
    the
  operators in the effective
  theory include light quark loops in general.
  Let us consider one of these diagrams:
  precisely the penguin diagram in Fig.\ 2, without QCD corrections. There is
  a light quark ($u$-quark or $c$-quark) running in the loop. We consider the
  zero mass limit for this quark.

  The pinch SP corresponding to the diagram in Fig.\ 2 is associated with a
  reduced diagram that coincides with the original one. At the pinch SP, the
  reduced diagram can be interpreted as a process occurring in space-time,
  with all internal particles real, on the mass-shell, and moving forward in
  time. Then it is easy to see that the two light quarks and the photon belong
  to the same jet, where a jet is defined as a connected set of massless
  lines, which are on shell with finite energy, and have momenta proportional
  to a single lightlike momentum ($p^\mu_\gamma$ in this case). Therefore, the
  reduced diagram in Fig.\ 2 can be viewed as a massive $b$ quark decaying
  into two
  jets, one consisting of the light quark loop and the photon, the other
  consisting of the $s$-quark line only. We use the term ``hard" for any
  vertex of a reduced diagram where lines from two or more jets are attached.
  We will also refer to on-shell massless lines with zero 4-momentum as soft
  lines; a ``soft subdiagram" is one consisting of only soft lines.

  Let us first analyze arbitrary reduced diagrams that contribute to the
  four-quark operator with two jets, $J_\gamma$ and $J_s$, a single hard part
  and a soft subdiagram.  Fig.\ 3 shows a typical diagrammatic structure
  for $J_\gamma$; Fig.\ 4 shows the general form of these
  diagrams. Afterwards, we shall treat 
  the remaining, relevant reduced diagrams. The diagrams of Fig.\ 4 admit a rather
  simple IR power counting, analogous to the treatment of form factors for
  quark-antiquark production in ${\rm e}^+{\rm e}^-$ annihilation \cite
  {sterman,stbook}. An appropriate choice of normal variables for all these
processes
  is the four components $k_s^\mu,\ \mu=0\dots 3$, of loops $k_s$ that pass
  through the soft subdiagram, $S$, and the {\em squares} $k_j^2$ and 
  $(k_{s,\perp}^j){}^2$ for each loop $k_j$ that is internal to a jet. For the
  latter, the transverse momentum is defined relative to the jet's direction.

  The superficial IR degree of divergence related to any reduced diagram of
  this sort is
  \begin{equation}
  D =\sum_{i=\{\gamma ,s\}}\,\left( 2L_i-N_i+b_i+\frac
  32f_i+t_i\right) ,
\label{pceq}
  \end{equation}
  where $L_i$ and $N_i$ are the numbers of loops and lines in $J_i$, while 
  $b_i$ and $f_i$ are the numbers of soft gluons and soft photons attached to 
  $J_i$ at the pinch SP. The factor $t_i$ comes from the numerator momenta in 
  $J_i$ and is bounded from below\footnote{For this
  argument, we ignore unphysically-polarized gluon lines that attach the jet
  to the hard scattering in covariant gauges; they do not affect the
overall power counting
  discussed here {\protect \cite{sterman,stbook}}.},
  \begin{equation}
  t_i\ge {\rm max}\left\{ \frac 12[u_3^i-v_i],0\right\} ,
\label{tinequal}
  \end{equation}
  where $u_3^i$
  and $v_i$ are, respectively,
   the number of three-point vertices in $J_i$
  and the
  number of soft vector particles attached to $J_i$. The Euler identity and counting relations
  between the numbers of lines and vertices of various orders may then be used
  to bound the IR degree of divergence by
  \begin{equation}
  D \ge \sum_{\{i=\gamma ,s\}}\left\{ {\frac 12}(h_i-1)+f_i+%
  {\frac 12}(v_i-u_3^i)\theta (v_i-u_3^i)\right\} \,,
  \label{pcinequal}
  \end{equation}
  where $h_i$ is the number of lines attaching jet $i$ to the decay vertex.
  Clearly, the lower limit on $D $ is found by taking $h_i=1$ and $f_i=0$.

  We can easily check that for the lowest-order diagram, Fig.\ 2, $D >0$,
  because in this case $h_\gamma =2$ for the photon jet, which immediately
  gives $D =1/2$. We may trace the positive value of $D $ to $u^\gamma_3=1
  $ in Eq.\ (\ref{tinequal}). This suppression is a direct result of the
  transversality of the emitted photon, which requires
  at least one power of the transverse momentum of the soft quark loop.
  Therefore, the diagram of Fig.\ 2 is IR convergent,
when the photon is on shell.  The all-order power counting
  expression, Eq.\ (\ref{pcinequal}) shows that this reasoning holds to any
  order for diagrams of the form of Fig.\ 4, since $h_\gamma =2$ for all of
  them when the hard vertex is a four-quark operator.

  We can readily extend this reasoning to the remaining pinch SP's that are
  relevant to $b\rightarrow s\gamma $. Further corrections to the four-quark
  operators are shown in Fig.\ 5. The only singular points involving an
  on-shell light quark penguin loop that we have not yet considered are those in which
  soft gluons attach to the $b$ quark (Fig.\ 5a) and those in which the 
penguin loop itself is soft (Fig.\ 5b).

  First, consider additional gluons attached to the $b$ quark. 
 The propagator of a quark radiating a soft gluon 
  behaves as $1/2 p \cdot k$
  (where $k$ is the momentum of the gluon) and so contributes $-1$ to the power
   counting.  For power counting purposes, the $b$ quark
  then acts like a third jet in Eq.\ \ref{pceq}, with no internal
  loops ($L_i=0$), no soft external fermions ($f_i=0$), no
  numerator suppression ($t_i=0$), and with the number of its lines
  equal to the number of soft gluons attached to the $b$ line
  ($N_i=b_i$).  It is easy to verify that pinch SP's of this sort
  leave Eq.\ (\ref{pcinequal}) unchanged.

  Second, consider the class of pinch SP's for
  the four-quark operator in which the photon is radiated by the $s$ (or $b$)
  quark, and the light quark penguin loop appears as part of the soft subdiagram
(Fig.\ 5b).
  This diagram is highly suppressed in the IR compared to those
  of Fig.\ 5a, because a fermion propagator with momentum $k^\mu$ diverges only linearly
  at $k^\mu=0$, in contrast to a boson propagator, which diverges
  quadratically.   Thus, penguin loops, whether connected directly
to the photon or not, are finite in the zero-mass limit.  

So far, we have restricted ourselves to light quarks in penguin
loops only, connected directly to the operators of the effective theory. 
Mixing of operators in the effective theory, however, 
gives rise to diagrams in which there is no penguin loop.
We will now show that in the two-body amplitude there are
no additional logarithms of light quark masses associated with
loops in the photon jet, that is, collinear to the outgoing photon.

 Consider  Fig.\ 6a, in which the gluon emerges from the
  hard subdiagram (effectively, the operator
  $O_8\sim m_b\,\bar s_L \sigma^{\mu\nu} T_a \, b_R \,G^a_{\mu\nu}$
  in the standard classification). These diagrams contain yet another
  set of pinch SP's, as shown, in which this gluon changes into a
  photon due to rescattering of a loop of
  virtual light quarks with soft gluons.
We may once again use the power
  counting of Eq.\ (\ref{pcinequal}), but this time $h_\gamma =1$, and on a
  diagram-by-diagram basis, the amplitude produces logarithms of the light
  quark mass. Note that charge conjugation invariance requires at least two
  soft gluons attached to the quark loop (two C-odd gluons cannot produce a
  C-odd photon), so that this effect appears first at three loops in the
  perturbative matrix element.
		Nevertheless, the contribution from any SP of this sort cancels in a gauge
  invariant set of diagrams, because the photon does not carry color. This
  result follows from the factorization of soft gluons from jets, an important
  ingredient in factorization proofs for inclusive cross sections \cite
  {CoSt81,CSSrv}.  

   Leaving technicalities aside, soft gluons can couple only to the total color
  charge of a jet. In fact, the contribution of the SP of Fig.\ 6a may also be
  pictured as in Fig.\ 6b, which shows that the total effect of the soft
  gluons is to insert a nonabelian phase factor in the color product between
  the decay vertex and the gluon line. The double line represents the
  nonabelian phase; the relevant Feynman rules are described, for example, in
  \cite{CSSrv}. Here, however, it is only the topology of the figure that is
  important. The remaining jet in Fig.\ 6b consists of a gluon and a photon,
  connected by the light quark loop, whose color trace vanishes identically.  
   Let us note that this factorization does not require
  a sum over final states; indeed, the cancellation of soft gluons
  in inclusive processes in Refs.\ \cite{CoSt81,CSSrv} requires the
  factorization as a first step, for each final state individually.
Thus, logarithms of $m_{loop}$
associated with these processes are absent, since, as we  have
just seen, a gluon cannot
fragment by interacting only with soft quanta.
Note that this reasoning does not
apply to final states 
describing the collinear splitting of  
the gluon into a photon plus other
gluons.  

In summary, we have shown, to all orders in $\alpha_s$ and to the lowest order 
in $\alpha_{ew}$, that penguin-like diagrams relevant for the process 
$b \rightarrow s (d) \gamma$ can be safely calculated by taking the 
massless limit for the $u$-quark circulating in the penguin loop.
This implies, for instance, that no large ``enhancements''
(i. e. IR unsafe contributions) in the amplitudes 
involving the CKM matrix elements $V_{ub} V^\star_{ud}$
are expected from this source at any fixed order in perturbation theory.
We have also observed that in the perturbative expansion of the two-body
final state, no logarithms of light-quark masses arise from 
loops collinear to the outoing photon.

\subsection*{Acknowledgements}

We wish to thank Howard Georgi,
Tobias Hurth, Michelangelo Mangano and A.\ Soni for helpful
conversations. We would also like to thank Z.\ Ligeti and J.\ Soares
for useful communications.
G.R.\ thanks the Institute for Theoretical Physics at
Stony Brook for its hospitality.
This work was supported in part by the National Science Foundation,
under grants PHY9722101 and PHY9218167.

\subsection*{Figure Captions}

\begin{itemize}
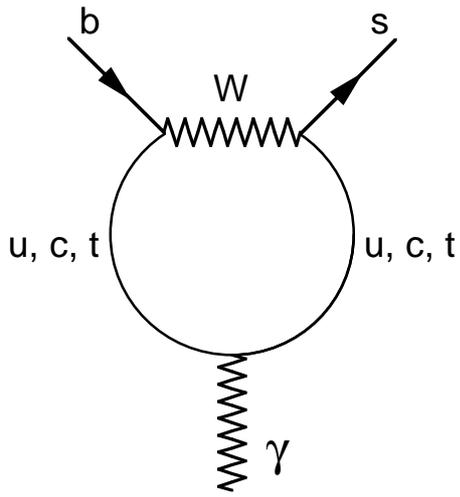
\item[Fig. 1 ]
One of the penguin diagrams
of the full theory, as computed, for example, in Ref.\ \cite{InamiLim}.
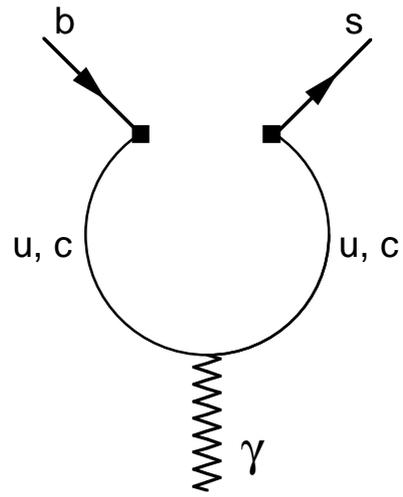
\item[Fig. 2 ]
 Penguin diagram of the effective (low energy) theory.
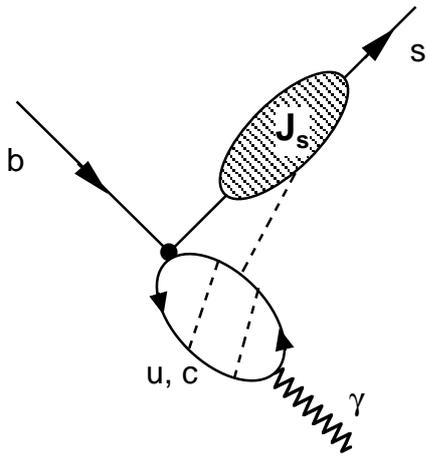
\item[Fig. 3 ]
Typical reduced diagram for the QCD corrections to the penguin-like diagrams.
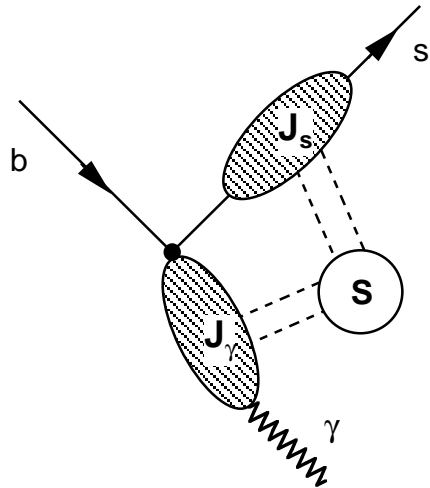
\item[Fig. 4 ]
General reduced diagram for soft gluon corrections to the  $ b \rightarrow
2$  jets transition.
\item[Fig. 5a]
Same as Fig. 4, but with gluonic insertions in the $b$-quark line.
\item[Fig. 5b]
Reduced diagram, analogous to those of Figs. 4, 5a, but with the $\gamma$
arising from an external ($b$ or $s$)  quark line, instead of the light
quark loop.
\item[Fig. 6a]
Reduced diagram with a gluon arising from the hard vertex.
The subdiagram hard gluon $\rightarrow$ soft gluons + $ \gamma $ vanishes
by color conservation
(see text).
\item[Fig. 6b]
Factorization of the soft gluon contribution to the diagram of Fig. 6a.
The double line represents the nonabelian phase referred to in the text.

\end{itemize}
  \end{document}